\documentstyle[12pt]{article}
\begin{document}
\begin{titlepage}
\begin{flushright}
IC/2001/108\\
hep-th/0108237
\end{flushright}
\vspace{10 mm}

\begin{center}
{\Large A Varying-$\alpha$ Brane World Cosmology}

\vspace{5mm}

\end{center}
\vspace{5 mm}

\begin{center}
{\large Donam Youm\footnote{E-mail: youmd@ictp.trieste.it}}

\vspace{3mm}

ICTP, Strada Costiera 11, 34014 Trieste, Italy

\end{center}

\vspace{1cm}

\begin{center}
{\large Abstract}
\end{center}

\noindent

We study the brane world cosmology in the RS2 model where the electric 
charge varies with time in the manner described by the varying 
fine-structure constant theory of Bekenstein.  We map such varying 
electric charge cosmology to the dual variable-speed-of-light cosmology 
by changing system of units.  We comment on cosmological implications 
for such cosmological models.  

\vspace{1cm}
\begin{flushleft}
August, 2001
\end{flushleft}
\end{titlepage}
\newpage

Recent study of quasar absorption line spectra in comparison with laboratory 
spectra has provided evidence that the fine structure constant $\alpha=
e^2/(4\pi\hbar c)$ varies over cosmological time scales \cite{wfc,mur,web}.  
If such result is true, then radical modification of standard physics may be 
required, because time-varying $\alpha$ implies time-varying fundamental 
constants of nature, i.e., the electric charge $e$, the Planck constant 
$\hbar$ and the speed of light $c$.  Bekenstein \cite{bek} constructed a 
varying-$e$ theory which preserves the Lorentz and the local gauge invariance 
but does not conserve electric charge.  Alternatively, Variable-Speed-of-Light 
(VSL) theories \cite{mof1,am} regard time variation of $\alpha$ as being due 
to time-varying $c$ and $\hbar$.  Since the speed of light varies with time, 
the Lorentz invariance is violated in VSL models.  The VSL cosmological 
models attracted some attention, because they resolve 
\cite{mof1,am,bar1,bar2,bar3,mof2,cm1,bar4,cm2,blm,cm3,cm4} 
the cosmological problems that are resolved by the inflationary models 
\cite{gut,lin,als}, along with some other cosmological problems.  

In our previous papers \cite{youm,youm1,youm2}, we studied VSL cosmologies in 
the Randall-Sundrum (RS) models \cite{rs1,rs2}, elaborating the resolution 
of cosmological problems.  The VSL models in the brane world scenarios are 
of interest, also because of the recent works 
\cite{kal1,kir,kal2,chu,ale,ish,ckr,csa} indicating the Lorentz violation 
in the brane world scenarios and because the VSL models may provide a 
mechanism for bringing the quantum correction to the fine-tuned brane tension 
under control.  In this paper, we study a varying-$\alpha$ cosmology in the 
RS model with one positive tension brane and infinite extra spatial dimension, 
i.e., the RS2 model \cite{rs2}, assuming time-varying electric charge $e$.  
(Cf. The standard cosmology with varying $e$ was previously  studied in Ref. 
\cite{sbm}.)  We briefly discuss the varying-$\alpha$ theory of Bekenstein 
\cite{bek} and incorporate the Bekenstein's theory into the brane world 
cosmology.  Then, we redefine system of units to map the varying-$e$ brane 
world cosmology to the dual VSL brane world cosmology.  We find that 
generally such dual VSL cosmological model cannot be used for resolving the 
cosmological problems.  However, our varying-$\alpha$ brane world 
cosmological models may be used for explaining the observed variation of 
$\alpha$ with time.  

We begin by discussing the varying-$\alpha$ theory of Bekenstein \cite{bek}.  
The Bekenstein's theory attributes the variation of the fine-structure 
constant $\alpha$ to the variation of electric charge $e$.  From the 
requirement that the charges of all particle species vary in exactly the 
same way (so that, for example, atoms can remain always neutral), we have 
$e=e_0\epsilon(x^{\mu})$, where $e_0$ is a constant characteristic of a 
particle species and $\epsilon$ is a dimensionless universal field.  Since 
$e$ is the electromagnetic coupling, the field $\epsilon$ is coupled to the 
gauge field as $\epsilon A_{\mu}$.  The appropriate gauge transformation law 
for $A_{\mu}$ is
\begin{equation}
\epsilon A_{\mu}\to\epsilon A_{\mu}+\partial_{\mu}\chi, 
\label{gaugtran}
\end{equation}
where $\chi$ is an arbitrary function of $x^{\mu}$.  The physical field 
strength $F_{\mu\nu}$, invariant under the gauge transformation 
(\ref{gaugtran}) and rescaling of $\epsilon$ by a constant factor, is given by
\begin{equation}
F_{\mu\nu}=\epsilon^{-1}\left[\partial_{\mu}(\epsilon A_{\nu})-
\partial_{\nu}(\epsilon A_{\mu})\right].
\label{fldstngth}
\end{equation}
This expression for $F_{\mu\nu}$ reduces to the usual form when $\epsilon$ 
is a constant.  The general electromagnetic action with varying $e$, 
satisfying the conditions of $(i)$ the Maxwellian electromagnetism and 
the minimal coupling of the vector potential to matter in the limit of 
constant $e$, $(ii)$ locally gauge invariant action, and $(iii)$ the 
time reversal invariance, is given by
\begin{equation}
S_{\rm em}=\int dx^4\sqrt{-g}\,{\cal L}_{\rm em}=-{1\over 4}\int dx^4\sqrt{-g}
\,F_{\mu\nu}F^{\mu\nu}.
\label{emact}
\end{equation}
The requirements that the variation of $e$ should result from dynamics and 
be derivable from an invariant action necessitate the introduction of a 
separate action for $\epsilon$.  The action for $\epsilon$, invariant under 
the rescaling of $\epsilon$ by a constant factor and giving rise to causal 
electromagnetism, is
\begin{equation}
S_{\epsilon}=-{1\over 2}{{\hbar c}\over l^2}\int d^4x\sqrt{-g}\,
{{\partial_{\mu}\epsilon\partial^{\mu}\epsilon}\over\epsilon^2},
\label{epsact}
\end{equation}
where $l$ is the length scale of the theory.  To simplify the action, we 
redefine the field $\epsilon$ as $\psi\equiv\ln\epsilon$.  Then, the action 
(\ref{epsact}) becomes
\begin{equation}
S_{\psi}=\int d^4\sqrt{-g}\,{\cal L}_{\psi}=-{\omega\over 2}\int d^4\sqrt{-g}
\,\partial_{\mu}\psi\partial^{\mu}\psi,
\label{psiact}
\end{equation}
where $\omega$ is a coupling constant.  

In this paper, we study the brane world cosmology in the RS2 model 
\cite{rs2}, incorporating the varying-$\alpha$ theory of Bekenstein.  
Since it is generally assumed in the brane world scenarios that photons 
are confined on the brane, we include the action (\ref{emact}) for the 
$U(1)$ gauge field $A_{\mu}$ in the brane action.  The action for the RS2 
model cosmology with varying-$e$ therefore takes the form
\begin{equation}
S=\int d^5x\sqrt{-G}\left[{c^4\over{16\pi G_5}}{\cal R}-\Lambda\right]+
\int d^4x\sqrt{-g}\left[{\cal L}_{\rm mat}+{\cal L}_{\psi}+e^{-2\psi}
{\cal L}_{\rm em}-\sigma\right],
\label{totact}
\end{equation}
where ${\cal L}_{\rm mat}$ is the Lagrangian for the matter fields on the 
brane, $G_5$ is the five-dimensional Newton's constant, $\Lambda$ is the 
bulk cosmological constant, and $\sigma$ is the tension of the brane assumed 
to be located at the origin $y=0$ of the extra spatial coordinate $y$.  Here, 
the metric $g_{\mu\nu}$ on the brane is given in terms of the bulk metric 
$G_{MN}$ by $g_{\mu\nu}(x^{\rho})=G_{\mu\nu}(x^{\rho},0)$.

Varying the action w.r.t. the metric, we obtain the Einstein's equations
\begin{equation}
{\cal G}_{MN}={{8\pi G_5}\over c^4}{\cal T}_{MN},
\label{eineqs}
\end{equation}
with the total energy-momentum tensor given by
\begin{equation}
{\cal T}_{MN}=-\Lambda G_{MN}+\delta^{\mu}_M\delta^{\nu}_N\left[
T^{\rm mat}_{\mu\nu}+T^{\psi}_{\mu\nu}+e^{-2\psi}T^{\rm em}_{\mu\nu}-
\sigma g_{\mu\nu}\right]{\sqrt{-g}\over\sqrt{-G}}\delta(y),
\label{emtens}
\end{equation}
where $T^{\rm mat}_{\mu\nu}=-{2\over\sqrt{-g}}{{\delta(\sqrt{-g}{\cal L}_{\rm 
mat})}\over{\delta g^{\mu\nu}}}$ is the energy-momentum tensor for the 
brane matter fields, and $T^{\psi}_{\mu\nu}$ and $T^{\rm em}_{\mu\nu}$ are 
respectively the energy-momentum tensors for the field $\psi$ and the $U(1)$ 
gauge field given by
\begin{eqnarray}
T^{\psi}_{\mu\nu}&=&\omega\partial_{\mu}\psi\partial_{\nu}\psi-{\omega\over 2}
g_{\mu\nu}\partial_{\rho}\psi\partial^{\rho}\psi,
\cr
T^{\rm em}_{\mu\nu}&=&F_{\mu\rho}F^{\ \,\rho}_{\nu}-{1\over 4}g_{\mu\nu}
F_{\alpha\beta}F^{\alpha\beta}.
\label{tens}
\end{eqnarray}
The following equation of motion for $\psi$ is obtained by varying the action 
w.r.t. $\psi$:
\begin{equation}
{\omega\over\sqrt{-g}}\partial_{\mu}\left[\sqrt{-g}g^{\mu\nu}\partial_{\nu}
\psi\right]=2e^{-2\psi}{\cal L}_{\rm em}.
\label{psieq}
\end{equation}

We consider the expanding brane universe where the principles of homogeneity 
and isotropy in the three-dimensional space on the three-brane are 
satisfied.  The general bulk metric ansatz is given by
\begin{equation}
G_{MN}dx^Mdx^N=-n^2(t,y)c^2dt^2+a^2(t,y)\gamma_{ij}dx^idx^j+b^2(t,y)dy^2,
\label{bulkmet}
\end{equation}
where $\gamma_{ij}$ is the metric for the maximally symmetric 
three-dimensional space given by
\begin{equation}
\gamma_{ij}dx^idx^j=\left(1+\textstyle{k\over 4}\delta_{mn}x^mx^n\right)^{-2}
\delta_{ij}dx^idx^j={{dr^2}\over{1-kr^2}}+r^2(d\theta^2+\sin^2\theta d\phi^2),
\label{threemet}
\end{equation}
with $k=-1,0,1$ respectively for the three-dimensional spaces with the 
negative, zero and positive spatial curvatures.  We define the time coordinate 
such that $n(t,0)=1$, i.e., $t$ is the cosmic time on the three-brane.  
The metric on the brane therefore takes the form
\begin{equation}
g_{\mu\nu}dx^{\mu}dx^{\nu}=-c^2dt^2+a^2_0(t)\gamma_{ij}dx^idx^j,
\label{branmet}
\end{equation}
where the subscript $0$ denotes the quantity evaluated at $y=0$, i.e., 
$a_0(t)\equiv a(t,0)$.  With the assumption of homogeneity and isotropy on 
the three-brane, the field $\psi$ does not depend on the spatial coordinates 
$x^i$ ($i=1,2,3$) of the three brane.  We substitute the above ansatz for 
fields into the equations of motion (\ref{eineqs},\ref{psieq}) to obtain 
the equations governing the evolution of the brane universe.

First, we consider the equation of motion (\ref{psieq}) for the field $\psi$.  
This equation determines how $e$, and therefore $\alpha$, varies with time.  
For the pure radiation, ${\cal L}_{\rm em}\sim F_{\mu\nu}F^{\mu\nu}$ 
vanishes.  So, only the nonrelativistic matter, whose dominant electromagnetic 
part of energy is Coulombic in nature, contributes to the RHS of Eq. 
(\ref{psieq}).  Denoting $\zeta$ as the fraction of mass density $\varrho_m$ 
of matter in the form of the Coulomb energy, we can put Eq. (\ref{psieq}) 
into the following form:
\begin{equation}
\ddot{\psi}+3H\dot{\psi}={2\over\omega}e^{-2\psi}\zeta\varrho_m,
\label{psieq2}
\end{equation}
where $H=\dot{a}_0/a_0$ is the Hubble parameter and the overdot denotes 
derivative w.r.t. $t$.  This equation expresses that during the radiation 
epoch, for which $\zeta\approx 0$, the time-variation of $e$ is negligible 
and $e$ varies significantly during the matter epoch, for which the universe 
is dominated by nonrelativistic matter and therefore $\zeta$ is not negligible.
When the universe becomes dominated by vacuum energy and begins to accelerate, 
$e$ approaches constant value.  

Next, we consider the Einstein's equations (\ref{eineqs}).  Note, the total 
energy density of the electromagnetic field is sum of the Coulomb energy 
density $\zeta\varrho_m$ of the brane matter and the radiation energy density 
$\varrho_r$.  So, the Einstein's equations reduce to
\begin{eqnarray}
{3\over{c^2n^2}}{\dot{a}\over a}\left({\dot{a}\over a}+{\dot{b}\over b}
\right)-{3\over b^2}\left[{a^{\prime}\over a}\left({a^{\prime}\over a}
-{b^{\prime}\over b}\right)+{a^{\prime\prime}\over a}\right]+{3k\over a^2}=
\ \ \ \ \ \ \ \ \ \ \ \ \ \ \ \ \ \ 
\cr
{{8\pi G_5}\over c^4}\left[\Lambda+\left\{\sigma+\varrho_mc^2(1+\zeta 
e^{-2\psi})+\varrho_rc^2e^{-2\psi}+\varrho_{\psi}c^2\right\}{{\delta(y)}
\over b}\right],
\label{ein1}
\end{eqnarray}
\begin{eqnarray}
{1\over b^2}\left[{a^{\prime}\over a}\left(2{n^{\prime}\over n}+{a^{\prime}
\over a}\right)-{b^{\prime}\over b}\left({n^{\prime}\over n}+2{a^{\prime}
\over a}\right)+2{a^{\prime\prime}\over a}+{n^{\prime\prime}\over n}\right]
\ \ \ \ \ \ \ \ \ \ \ \ \ \ \ \ \ \ \ \ \ \ 
\cr
+{1\over{c^2n^2}}\left[{\dot{a}\over a}\left(2{\dot{n}\over n}-{\dot{a}\over 
a}\right)+{\dot{b}\over b}\left({\dot{n}\over n}-2{\dot{a}\over a}\right)-2
{\ddot{a}\over a}-{\ddot{b}\over b}\right]-{k\over a^2}=\ \ \ \ \ 
\cr
{{8\pi G_5}\over c^4}\left[-\Lambda+\left(\textstyle{1\over 3}\varrho_r
e^{-2\psi}+\varrho_{\psi}-\sigma\right){{\delta(y)}\over b}\right],
\label{ein2}
\end{eqnarray}
\begin{equation}
{n^{\prime}\over n}{\dot{a}\over a}+{a^{\prime}\over a}{\dot{b}\over b}
-{\dot{a}^{\prime}\over a}=0,
\label{ein3}
\end{equation}
\begin{equation}
{3\over b^2}{a^{\prime}\over a}\left({a^{\prime}\over a}+{n^{\prime}\over n}
\right)-{3\over{c^2n^2}}\left[{\dot{a}\over a}\left({\dot{a}\over a}-{\dot{n}
\over n}\right)+{\ddot{a}\over a}\right]-{{3k}\over a^2}=-{{8\pi G_5}\over 
c^4}\Lambda,
\label{ein4}
\end{equation}
where prime denotes derivative w.r.t. $y$ and $\varrho_{\psi}={\omega\over 2}
{\dot{\psi}^2\over c^4}$ is the mass density of the field $\psi$.

Due to the $\delta$-function singularity at $y=0$, the first derivatives 
of the metric components $a$ and $n$ are discontinuous at $y=0$.  From 
Eqs. (\ref{ein1},\ref{ein2}), we obtain the following boundary conditions 
on $a^{\prime}$ and $n^{\prime}$ at $y=0$:
\begin{equation}
{{[a^{\prime}]_0}\over{a_0b_0}}=-{{8\pi G_5}\over{3c^4}}\left[\sigma+
\varrho_mc^2(1+\zeta e^{-2\psi})+\varrho_rc^2e^{-2\psi}+\varrho_{\psi}c^2
\right],
\label{bc1}
\end{equation}
\begin{equation}
{{[n^{\prime}]_0}\over{n_0b_0}}=-{{8\pi G_5}\over{3c^4}}\left[\sigma-2
\varrho_mc^2(1+\zeta e^{-2\psi})-3\varrho_rc^2e^{-2\psi}-5\varrho_{\psi}c^2
\right],
\label{bc2}
\end{equation}
where $[F]_0\equiv F(0^+)-F(0^-)$ denotes the jump of $F(y)$ across $y=0$.

We now obtain the effective Friedmann equations for the four-dimensional 
universe on the brane worldvolume, following the procedure developed in 
Ref. \cite{bdl}.  The effective four-dimensional equations on the brane 
worldvolume are obtained by taking the jumps and the mean values of the 
above five-dimensional equations of motion across $y=0$, applying the 
boundary conditions (\ref{bc1},\ref{bc2}) on the first derivatives.  
Here, the mean value of a function $F$ across $y=0$ is defined as $\sharp 
F\sharp\equiv[F(0^+)+F(0^-)]/2$.  We consider the solution invariant under 
the ${\bf Z}_2$ symmetry, $y\to -y$.  

First, by taking the jump of the $(t,y)$-component Einstein's equation 
(\ref{ein3}), we obtain the following conservation equation:
\begin{equation}
\dot{\varrho}+3\left(\varrho+{\wp\over c^2}\right){\dot{a}_0\over a_0}=0,
\label{cnseq}
\end{equation}
where $\varrho$ and $\wp$ are total mass density and pressure of fields on 
the brane given by
\begin{equation}
\varrho=\varrho_m(1+\zeta e^{-2\psi})+\varrho_re^{-2\psi}+\varrho_{\psi},
\ \ \ \ \ \ \ \ \ \ \ \ 
\wp=\textstyle{1\over 3}\varrho_re^{-2\psi}c^2+\varrho_{\psi}c^2.
\label{denspres}
\end{equation}
From the mean values of the diagonal component Einstein equations, we 
obtain the following effective Friedmann equations on the brane worldvolume:
\begin{equation}
\left({\dot{a}_0\over a_0}\right)^2={{16\pi^2G^2_5}\over{9c^6}}(\varrho^2c^4
+2\sigma\varrho c^2)+{{{\cal C}c^2}\over a^4_0}+{{4\pi G_5}\over{3c^2}}
\left(\Lambda+{{4\pi G_5}\over{3c^4}}\sigma^2\right)-{{kc^2}\over a^2_0},
\label{frd1}
\end{equation}
\begin{equation}
{\ddot{a}_0\over a_0}=-{{16\pi^2G^2_5}\over{9c^6}}(2\varrho^2c^4+\sigma
\varrho c^2+3\sigma\wp+3\wp\varrho c^2)-{{{\cal C}c^2}\over a^4_0}+{{4\pi 
G_5}\over{3c^2}}\left(\Lambda+{{4\pi G_5}\over{3c^4}}\sigma^2\right),
\label{frd2}
\end{equation}
where ${\cal C}$ is an integration constant.  In obtaining the equations, 
it is assumed that the radius of the extra space is stabilized, i.e., 
$\dot{b}=0$, and the $y$-coordinate is defined to be proportional to the 
proper distance along the $y$-direction with $b$ being the constant of 
proportionality, i.e., $b^{\prime}=0$.  Using these assumptions, we have 
defined the $y$-coordinate such that $b=1$.  

As was pointed out in Ref. \cite{am}, physically it does not make sense 
to talk about constancy or variability of dimensional constants such 
as the speed of light, since the measured values of any dimensional 
quantities are actually the ratio to some standard units, which may vary 
with time.  So, strictly we can experimentally test the constancy of 
dimensionless ratios of dimensional quantities, only.  For a given 
dimensionless ratio of dimensional quantities, which is experimentally 
observed to vary with time, we can arbitrarily choose units such that any 
group of dimensional quantities (forming the ratio) vary with time and the 
remaining ones do not.  For example, we can regard the time variation of the 
(dimensionless) fine-structure constant $\alpha=e^2/(4\pi\hbar c)$ as being 
due to the time variation of $e$, as was considered by Bekenstein \cite{bek}, 
or due to the time variation of $c$ and $\hbar$, as in the VSL models, 
depening on the choice of units.  Furthermore, we can always redefine system 
of units such that a given model is mapped to the dual model for which 
different group of dimensional quantities vary with time.  In the following, 
we study the VSL dual of the brane world cosmology with time-varying $e$ 
considered in the above.  

We redefine system of units such that the electric charge remains constant 
and speed of light varies with time.  The measured quantities in the system 
of units where the speed of light takes constant value $c$ are denoted 
without hats.  Those in the system of units in which the speed of light 
$\hat{c}$ varies with time (with the electric charge remaining constant) 
are denoted with hat.   We define the new system of units such that 
measurements in the two systems of units are related as (Cf. Ref. \cite{bm})
\begin{equation}
c^2dt=\hat{c}^2d\hat{t},\ \ \ \ \ \ \ \ \ \ 
cdx=\hat{c}d\hat{x},\ \ \ \ \ \ \ \ \ \ 
{{dE}\over c^3}={{d\hat{E}}\over\hat{c}^3}.
\label{msrmntrel}
\end{equation}
To find the relations between dimensional quantities in the two systems of 
units, we consider the following ratios of the dimensional quantities:
\begin{equation}
{{cdt}\over{dx}}={{\hat{c}d\hat{t}}\over{d\hat{x}}},\ \ \ \ \ \ \ 
{\hbar\over{dEdt}}={\hat{\hbar}\over{d\hat{E}d\hat{t}}},\ \ \ \ \ \ \ 
{{G_4dE}\over{c^4dx}}={{\hat{G}_4d\hat{E}}\over{\hat{c}^4d\hat{x}}},
\ \ \ \ \ \ \ 
{e^2\over{dEdx}}={\hat{e}^2\over{d\hat{E}d\hat{x}}},
\label{dmlssrts}
\end{equation}
where $G_4=4\pi G^2_5\sigma/(3c^4)$ is the effective four-dimensional 
Newton's constant, which can be read off from Eqs. (\ref{frd1},\ref{frd2}), 
and $\hat{e}=e_0=e/\epsilon={\rm const}$.  Note, the dimensionless ratios 
take the same values regardless of the choice of system of units.  
Substituting Eq. (\ref{msrmntrel}) into Eq. (\ref{dmlssrts}), we find that 
the speed of light and the Planck constant vary with time in the following 
manner and the effective Newton's constant remains constant:
\begin{equation}
\hat{c}=c/\epsilon,\ \ \ \ \ \ \ \ \ \ \ \ 
\hat{\hbar}=\hbar/\epsilon,\ \ \ \ \ \ \ \ \ \ \ \ 
\hat{G}_4=G_4.
\label{qntrels}
\end{equation}
The transformation (\ref{msrmntrel}) then reduces to
\begin{equation}
d\hat{t}=\epsilon^2dt,\ \ \ \ \ \ \ \ \ \ \ \ \ 
d\hat{x}=\epsilon dx,\ \ \ \ \ \ \ \ \ \ \ \ \ \ \ 
d\hat{E}=dE/\epsilon^3.
\label{trnsf}
\end{equation}
This transformation determines relation between measurements in the two 
systems of units.  In particular, the mass densities, the pressures and the 
brane tensions in the two systems of units are related as
\begin{equation}
\hat{\varrho}=\varrho/\epsilon^4,\ \ \ \ \ \ \ \ \ \ \ 
\hat{\wp}=\wp/\epsilon^6,\ \ \ \ \ \ \ \ \ \ \ 
\hat{\sigma}=\sigma/\epsilon^6.
\label{denprsrels}
\end{equation}
From Eqs. (\ref{qntrels},\ref{denprsrels}), we see that the five-dimensional 
Newton's constants in the two systems of units are related as $\hat{G}_5= 
G_5\epsilon$.  

In the new system of units, in which the speed of light $\hat{c}$ 
varies with time and the electric charge $\hat{e}=e_0$ remains 
constant, the bulk metric (\ref{bulkmet}) takes the form
\begin{equation}
d\hat{s}^2=\epsilon^2ds^2=-\hat{n}^2\hat{c}^2d\hat{t}^2+\hat{a}^2
\hat{\gamma}_{ij}d\hat{x}^id\hat{x}^j+\hat{b}^2d\hat{y}^2,
\label{newblkmet}
\end{equation}
where $\hat{n}=n$, $\hat{a}=\epsilon a$, $\hat{b}=b$, and 
\begin{equation}
\hat{\gamma}_{ij}d\hat{x}^id\hat{x}^j=\left(1+\textstyle{\hat{k}\over 
4}\delta_{mn}\hat{x}^m\hat{x}^n\right)\delta_{ij}d\hat{x}^id\hat{x}^j=
{{d\hat{r}^2}\over{1-\hat{k}\hat{r}^2}}+\hat{r}^2\left(d\hat{\theta}^2+
\sin^2\hat{\theta}d\hat{\phi}^2\right).
\label{hatgam}
\end{equation}
After the transformation of units, we performed a spatial coordinate 
transformation so that $\hat{k}=k=0,\pm 1$.  
So, the effective Friedmann equations (\ref{frd1},\ref{frd2}) in the new 
system of units take the forms
\begin{equation}
\left({\dot{\hat{a}_0}\over\hat{a}_0}+{\dot{\hat{c}}\over\hat{c}}\right)^2=
{{16\pi^2\hat{G}^2_5}\over{9\hat{c}^6}}\left(\hat{\varrho}^2\hat{c}^4+
2\hat{\sigma}\hat{\varrho}\hat{c}^2\right)+{{\hat{\cal C}\hat{c}^2}\over
\hat{a}^4_0}+{{4\pi\hat{G}_5}\over{3\hat{c}^2}}\left(\hat{\Lambda}+
{{4\pi\hat{G}_5}\over{3\hat{c}^4}}\hat{\sigma}^2\right)-{{\hat{k}
\hat{c}^2}\over\hat{a}^2_0},
\label{nfrdeq1}
\end{equation}
\begin{equation}
{\ddot{\hat{a}}_0\over\hat{a}_0}+{\ddot{\hat{c}}\over\hat{c}}-2{\dot{\hat{c}}^2
\over\hat{c}^2}=
-{{16\pi^2\hat{G}^2_5}\over{9\hat{c}^6}}\left(2\hat{\varrho}^2\hat{c}^4+
\hat{\sigma}\hat{\varrho}\hat{c}^2+3\hat{\sigma}\hat{\wp}+3\hat{\wp}
\hat{\varrho}\hat{c}^2\right)-{{\hat{\cal C}\hat{c}^2}\over\hat{a}^4_0}
+{{4\pi\hat{G}_4}\over{3\hat{c}^2}}\left(\hat{\Lambda}+{{4\pi\hat{G}_5}
\over{3\hat{c}^4}}\hat{\sigma}^2\right),
\label{nfrdeq2}
\end{equation}
where $\hat{\cal C}=\epsilon^2{\cal C}$, $\hat{\Lambda}=\Lambda/\epsilon$, 
the overdot from now on denotes derivative w.r.t. $\hat{t}$, and the total 
mass density and pressure are given by
\begin{eqnarray}
\hat{\varrho}&=&\hat{\varrho}_m\left[1+\zeta\left({\hat{c}\over c}\right)^2
\right]+\hat{\varrho}_r\left({\hat{c}\over c}\right)^2+{1\over 2}{\omega
\over c^4}{\dot{\hat{c}}^2\over\hat{c}^2},
\cr
{\hat{\wp}\over\hat{c}^2}&=&{1\over 3}\hat{\varrho}_r\left({\hat{c}\over 
c}\right)^2+{1\over 3}{\omega\over c^4}{\dot{\hat{c}}^2\over\hat{c}^2}.
\label{denprs}
\end{eqnarray}
We made use of the relation $\epsilon=c/\hat{c}$ to express $\epsilon$ 
in terms of the speed of light.

The equation of motion (\ref{psieq2}) for the field $\psi=\ln\epsilon=\ln 
(c/\hat{c})$ becomes the following equation that determines the temporal 
variation of the speed of light $\hat{c}$:
\begin{equation}
{\ddot{\hat{c}}\over\hat{c}}+3{\dot{\hat{a}}_0\over\hat{a}_0}{\dot{\hat{c}}
\over\hat{c}}=-{2\over\omega}\zeta\left({\hat{c}\over c}\right)^2
\hat{\varrho}_m.
\label{eqfrc}
\end{equation}
The temporal variation of $\hat{c}$ is driven by electrostatic Coulomb 
energy, parameterized by $\zeta$.  From this equation, we see that the 
temporal variation of the speed of light is negligible during the radiation 
dominated epoch (for which $\zeta\approx 0$), and the speed of light varies 
noticeably during the matter dominated epoch.  The amount of variation is 
determined by the ratio $\zeta/\omega$.  Once the universe accelerates, 
the speed of light approaches the constant present day value.

The conservation equation (\ref{cnseq}) becomes
\begin{equation}
\dot{\hat{\varrho}}+3\left(\hat{\varrho}+{\hat{\wp}\over\hat{c}^2}\right)
{\dot{\hat{a}}_0\over\hat{a}_0}=\left(\hat{\varrho}-3{\hat{\wp}\over
\hat{c}^2}\right){\dot{\hat{c}}\over\hat{c}}.
\label{nwcnsveq}
\end{equation}
Making use of the explicit expressions (\ref{denprs}) for $\hat{\varrho}$ 
and $\hat{\wp}$, we can reexpress the conservation equation as
\begin{eqnarray}
\dot{\hat{\varrho}}_{\rm fld}+3\left(\hat{\varrho}_{\rm fld}+
{\hat{\wp}_{\rm fld}\over\hat{c}^2}\right){\dot{\hat{a}}_0\over\hat{a}_0}=
\left[\hat{\varrho}_{\rm fld}-3{\hat{\wp}_{\rm fld}\over\hat{c}^2}+{\omega
\over c^4}\left({\dot{\hat{c}}^2\over\hat{c}^2}-{\ddot{\hat{c}}\over\hat{c}}
-{5\over 2}{\dot{\hat{c}}\over\hat{c}}{\dot{\hat{a}}_0\over\hat{a}_0}\right)
\right]{\dot{\hat{c}}\over\hat{c}}
\cr
=\left[\hat{\varrho}_{\rm fld}-3{\hat{\wp}_{\rm fld}\over\hat{c}^2}+{\omega
\over c^4}\left\{{\dot{\hat{c}}^2\over\hat{c}^2}-{1\over 6}{\ddot{\hat{c}}
\over\hat{c}}+{5\over 3}{\zeta\over\omega}\left({\hat{c}\over c}\right)^2
\hat{\varrho}_m\right\}\right]{\dot{\hat{c}}\over\hat{c}},
\label{csveq}
\end{eqnarray}
where we redefined mass density and pressure to contain contribution from 
the brane matter field and the electromagnetic field, only:
\begin{equation}
\hat{\varrho}_{\rm fld}=\hat{\varrho}_m\left[1+\zeta\left({\hat{c}\over c}
\right)^2\right]+\hat{\varrho}_r\left({\hat{c}\over c}\right)^2,\ \ \ \ \ \ \ 
\hat{\wp}_{\rm fld}={1\over 3}\hat{\varrho}_r\left({\hat{c}\over 
c}\right)^2,
\label{flddenprs}
\end{equation}
and on the second line of Eq. (\ref{csveq}) we made use of Eq. (\ref{eqfrc}).  
So, in the new system of units the total mass density $\hat{\varrho}_{\rm 
fld}$ of fields on the brane is not preserved while the speed of light 
$\hat{c}$ varies with time.  Unlike the case of the usual VSL models, the RHS 
of the conservation equation has no dependence on the curvature parameter 
$\hat{k}$.  This implies that for any circumstances $\hat{k}=0$ does not 
correspond to the stable attractor.  The matter will be produced or taken 
away while $\hat{c}$ varies with time, even when $\hat{k}=0$, i.e., when the 
mass density of the brane universe reached the critical density.  So, the VSL 
model, dual to the brane world cosmology with time-varying $e$, cannot by 
itself resolve the flatness problem.  

To make contact with conventional cosmology, we assume that $\hat{\sigma}\gg
\hat{\varrho}\hat{c}^2, \hat{\wp}$.  To the leading order, the Friedmann 
equations (\ref{nfrdeq1},\ref{nfrdeq2}) take the forms
\begin{equation}
\left({\dot{\hat{a}_0}\over\hat{a}_0}+{\dot{\hat{c}}\over\hat{c}}\right)^2=
{{8\pi\hat{G}_4}\over 3}\hat{\varrho}+{\hat{c}^2\over 3}\hat{\Lambda}_{\rm eff}
+{\hat{\cal C}\hat{c}^2\over\hat{a}^4_0}-{{\hat{k}\hat{c}^2}\over\hat{a}^2_0},
\label{apxfrd1}
\end{equation}
\begin{equation}
{\ddot{\hat{a}_0}\over\hat{a}_0}+{\ddot{\hat{c}}\over\hat{c}}-2{\dot{\hat{c}}^2
\over\hat{c}^2}=-{{4\pi\hat{G}_4}\over 3}\left(\hat{\varrho}+3{\hat{\wp}\over
\hat{c}^2}\right)+{\hat{c}^2\over 3}\hat{\Lambda}_{\rm eff}-{{\hat{\cal C}
\hat{c}^2}\over\hat{a}^4_0},
\label{apxfrd2}
\end{equation}
where $\hat{G}_4=4\pi\hat{G}^2_5\hat{\sigma}/(3\hat{c}^4)$ and 
$\hat{\Lambda}_{\rm eff}$ is the effective four-dimensional cosmological 
constant given by
\begin{equation}
\hat{\Lambda}_{\rm eff}={{4\pi\hat{G}_5}\over\hat{c}^4}\left(
\hat{\Lambda}+{{4\pi\hat{G}_5}\over{3\hat{c}^4}}\hat{\sigma}^2\right).
\label{efflamb}
\end{equation}
Note, in the new system of units, the bulk cosmological constant and the 
brane tension vary with time as $\hat{\Lambda}\sim 1/\epsilon$ and 
$\hat{\sigma}\sim 1/\epsilon^6$.  However, once the brane tension initially 
takes the fine-tuned value giving rise to $\Lambda_{\rm eff}=0$, $\Lambda_{\rm 
eff}$ will stay zero throughout the cosmological evolution as long as there is 
no other contribution giving rise to the correction to $\hat{\sigma}$.  

Making use of the explicit expressions (\ref{denprs}) for $\hat{\varrho}$ and 
$\hat{\wp}$, we can reexpress these Friedmann equations as
\begin{equation}
\left({\dot{\hat{a}}_0\over\hat{a}_0}+{\dot{\hat{c}}\over\hat{c}}\right)^2=
{{8\pi\hat{G}_4}\over 3}\hat{\varrho}_{\rm fld}+{{4\pi\omega\hat{G}_4}\over
{3c^4}}{\dot{\hat{c}}^2\over\hat{c}^2}+{\hat{c}^2\over 3}
\hat{\Lambda}_{\rm eff}+{\hat{\cal C}\hat{c}^2\over\hat{a}^4_0}-{{\hat{k}
\hat{c}^2}\over\hat{a}^2_0},
\label{efred1}
\end{equation}
\begin{equation}
{\ddot{\hat{a}}_0\over\hat{a}_0}+{\ddot{\hat{c}}\over\hat{c}}-2{\dot{\hat{c}}^2
\over\hat{c}^2}=-{{4\pi\hat{G}_4}\over 3}\left(\hat{\varrho}_{\rm fld}+3
{\hat{\wp}_{\rm fld}\over\hat{c}^2}\right)-{{2\pi\omega\hat{G}_4}\over c^4}
{\dot{\hat{c}}^2\over\hat{c}^2}+{\hat{c}^2\over 3}
\hat{\Lambda}_{\rm eff}-{{\hat{\cal C}\hat{c}^2}\over\hat{a}^4_0}.
\label{efred2}
\end{equation}
We notice that the effective Friedmann equations for the dual VSL cosmology 
have different structure from those of the VSL brane world cosmologies that 
we studied in our previous papers \cite{youm,youm1,youm2}.  Namely, there 
are additional terms involving the time derivatives of the speed of light 
$\hat{c}$.  Such additional terms arose due to the fact that the above 
effective Friedmann equations for the dual VSL cosmology do not describe 
expanding universe in the preferred frame, in which the equations of motion 
take the conventional forms with the constant speed of light just replaced 
by the variable speed of light.  In the preferred frame, the above effective 
Friedmann equations for the dual VSL cosmology would take the forms of the 
Friedmann equations for the ordinary VSL cosmologies.  Note, our dual VSL 
cosmology cannot by itself resolve the flatness problem, whereas the usual 
VSL brane world cosmologies \cite{youm,youm1,youm2} can.  So, whether a 
particular cosmological problem can be resolved or not depends on the frame 
used.  However, cosmological problems that require faster speed of light 
during very early stage of comological evolution such as the horizon problem 
and the problem of unwanted relics cannot be resolved by our varying-$\alpha$ 
cosmology regardless of the frame used, because during the very early stage 
of the comological evolution the universe is dominated by radiation and 
therefore the speed of light does not take large value (due to negligible 
$\zeta$).  Nevertheless, our varying-$\alpha$ cosmological model may be used 
for explaining time-varying fine-structure constant observed in our universe.

\end{document}